\documentclass[aps, preprint, groupedaddress, floatfix]{revtex4}
\usepackage{epsfig}
\usepackage{graphicx}
\usepackage{amsmath}

\begin{document}

\title{The Fundamental Asymmetry in Interfacial Electronic Reconstruction between Insulating Oxides}

\author{Hanghui Chen$^{1,3}$, Alexie M. Kolpak$^{2,3}$ and Sohrab Ismail-Beigi$^{1,2,3}$}

\affiliation{$^1$Department of Physics, Yale University, New
Haven, Connecticut, 06511 USA \\
$^2$Department of Applied Physics, Yale University, New
Haven,Connecticut, 06511 USA \\
$^3$Center for Research on Interface Structures and Phenomena
(CRISP), Yale University, New Haven, Connecticut, 06511 USA }

\date{\today}

\begin{abstract}
We present an \textit{ab initio} study of the (001) interfaces
between two insulating perovskites, the polar LaAlO$_3$ and the
non-polar SrTiO$_3$. We observe an insulating-to-metallic
transition above a critical LaAlO$_3$ thickness. We explain that
the high conductivity observed at the TiO$_2$/LaO interface, and
the lack of similar conductivity at the SrO/AlO$_2$ interface, are
inherent in the atomic geometry of the system. A large interfacial
hopping matrix element between cations causes the formation of a
bound electron state at the TiO$_2$/LaO interface. This novel
mechanism for the formation of interfacial bound states suggests a
robust means for tuning conductivities at various oxide
heterointerfaces.

\end{abstract}

\maketitle

A wide variety of novel physical phenomena are observed at the
hetero-epitaxial interfaces between perovskite oxides~\cite{1},
such as high conductivity~\cite{3,4}, external field
doping~\cite{5,6}, magnetism~\cite{7} and
supercoductivity~\cite{8}, all of which are lacking in the bulk
constituents. Understanding the rich physics of these systems, and
their potential applications, has been the focus of intensive
research in the past decade.

Recently, the discovery of many novel properties at the (001)
epitaxial interface LaAlO$_3$/SrTiO$_3$ (LAO/STO) has boosted
considerable research ~\cite{4,5,6,7,8,9,10,11,12}. Since both LAO
and STO are perovskite oxides, two types of interfaces can form
along the (001) direction: TiO$_2$/LaO ($n$-type) and SrO/AlO$_2$
($p$-type)~\cite{4,9}. While experiments have found that the
$p$-type interface is insulating, a high-mobility quasi
two-dimensional electron gas (Q2-DEG) is formed at the $n$-type
interface~\cite{4} when the LAO thickness exceeds a critical
separation~\cite{5,10}. The origin of the Q2-DEG is the key to
understanding the novelty of the LAO/STO interface, but it is
still under heated debate in both
theory~\cite{13,14,15,16,17,18,19,20,21} and
experiment~\cite{9,22,23,24,25}.

Broadly, there are two representative schools of thought. The
first suggests an intrinsic mechanism due to the polar
discontinuity at the interface~\cite{4,9}. Since the nominal
charges of the LaO and AlO$_2$ planes in (001) LAO are +1 and -1
(STO has neutral SrO and TiO$_2$ layers), there is a macroscropic
electric field through the LAO. The voltage across the LAO is
approximately proportional to its thickness and diverges with
increasing thickness. To avoid the divergence, half an
electron(hole) per two-dimensional unit cell must transfer across
the $n$-type($p$-type) interface, respectively~\cite{4,9}.
Nevertheless, this symmetric description can not account for the
drastic difference in electronic properties of the two interfaces.
The second school of thought posits that extrinsic oxygen
vacancies generated by the pulsed laser deposition create the
dominating carriers responsible for the
conductivity~\cite{9,22,23,24}. However, both the existence of a
lower limit in the carrier density of fully oxidized $n$-type
interfaces~\cite{25} and the insulating property of $p$-type
interfaces~\cite{4,9} strongly suggest that a more fundamental
intrinsic asymmetry exists in the electronic
reconstruction~\cite{26} between these two interfaces. In this
Letter we reveal a fundamental difference in the electronic
reconstruction between $p$-type and $n$-type interfaces and
propose that the atomic geometry accounts for the difference.

We begin by considering a schematic of the local energy gap in
Fig. 1. We start with the simplest $np$-type geometry in which
several LAO layers are sandwiched between two thick slabs of STO
with an $n$-type interface on the left and a $p$-type interface on
the right. In both STO slabs, the local energy gap equals that of
bulk STO (3.2 eV). For the LAO film it has the bulk LAO value (5.6
eV) except that the macroscopic electric field through the polar
LAO lifts up its local energy gap, as illustrated in Fig. 1a. The
energy gap of the entire interface system is thus determined by
the difference between the conduction band edge (CBE) of STO on
the left and the valence band edge (VBE) of STO on the right.
Therefore, the LAO film serves to diminish the gap. As the LAO
film thickens, the voltage across the LAO accrues until the VBE of
STO on the right exceeds the CBE of STO on the left (bottom panel
of Fig. 1a). At this point, the CBE on the left become more
energically favorable than the VBE on the right and electrons
transfer across the LAO. These transferred electrons reside on the
Ti $d$-orbitals of the CBE on the left while holes are left behind
on the O $p$-orbitals of the VBE on the right. This picture also
applies to the $n$-type interface (Fig. 1b) except that the holes
stay in the O $p$-orbitals of the LAO surface valence bands. For
the $p$-type interface (Fig. 1c), the polarity of the LAO film is
reversed so the local band edges bend down, and the electron
transfer occurs when the CBE of LAO is lower than the VBE of STO.
In all three cases, the condition for electron transfer is that
the voltage across the LAO film exceeds the band gap of STO
($np$-type and $n$-type) or the band gap of LAO ($p$-type). The
minimal LAO thickness satisfying this condition is the critical
separation.

To verify and move beyond schematics, we employ density functional
theory within the \textit{ab initio} supercell plane-wave
approach~\cite{27} with the local density approximation
(LDA)~\cite{28}, and ultrasoft pseudopotentials~\cite{29}. All the
super cells~\cite{33} in our simulation have the format
(STO)$_{s_1}$/(LAO)$_{l}$/(STO)$_{s_2}$/vacuum, in which $s_1,
s_2$ are the numbers of STO unit cell and $l$ is that of LAO.
$s_1, s_2$ could be half integers due to an extra atomic layer
(i.e. half a unit cell). The vacuum separating periodic copies is
$\sim 15$ \AA \phantom iin the (001) direction. For the $n$-type
and $p$-type interfaces, $s_2$ is set to zero. For the $np$-type,
the interface between (STO)$_{s_1}$ and (LAO)$_l$ is $n$-type and
that between (LAO)$_l$ and (STO)$_{s_2}$ is $p$-type. Typical
values of $s_1$ and $s_2$ are 5 (for $np$-type) and 11 (for
$n$-type and $p$-type); $l$ is varied from 1 to 7. The interfaces
are along (001) so the $z$ axis is perpendicular to the
interfaces. The lengths in $x$ and $y$ directions of the unit
cell, subject to periodic boundary conditions, are fixed to the
theoretical lattice constant of STO $a=3.85$\AA \phantom i(1.5\%
smaller than the experimental value). All atoms except two STO
layers facing the surface~\cite{30} are fully relaxed in the $z$
direction until every force component is smaller than 26 meV/\AA.
The charge densities in Fig. 2 have been checked with convergence
by reducing the force components to 13 meV/\AA \phantom{ } or
smaller.

Our \textit{ab initio} calculations support the above schematic
view. For $n$-type interfaces, once the transfer occurs, electrons
accumulate at the interface and inhabit electronic states composed
of Ti $d$-orbitals in the STO; for a (001) growth direction, the
states have clear $d_{xy}$ character. For $p$-type interfaces,
holes inhabit states of pure O $p$-character in the STO. Electron
transfer occurs at 3, 4 and 5 LAO unit cells (u.c.) within DFT-LDA
for $np$-type, $n$-type, and $p$-type interfaces, respectively.
Table 1 lists the energy gap $\Delta$ versus the number of LAO
u.c. as well as the corresponding potential change $V$ per added
LAO u.c. Based on the data, $V$ extrapolates to 0.7 eV.

This potential change is predicted accurately by DFT as it is a
ground-state property depending only on the electron density.
However, since DFT describes excitations poorly, the computed STO
and LAO band gaps are underestimated (1.85 and 3.25 eV,
respectively) relative to experiments. Therefore, we use the
experimental band gaps of STO and LAO to estimate the actual
critical separation. The difference between the experimental and
calculated band gaps is taken into account by adding extra LAO
layers, each of which reduces the energy gap by 0.7 eV. Thus we
predict experimental critical separations of 5, 6 and 8 LAO u.c.
for $np$-type, $n$-type and $p$-type interfaces,
respectively~\cite{31}. The critical separation for the $np$-type
interface agrees very well with experiment~\cite{10} while our
prediction for the $n$-type interface (6 u.c.) is larger than the
experimental value (4 u.c.)~\cite{5}. An important physical reason
might be that we simulate perfectly flat, periodic, and
defect-free LAO surfaces exposed to vacuum. In experiment,
however, unavoidable imperfections of the surfaces (e.g. due to
reconstructions or defects) may be responsible for an abrupt
transition in the conductivities at smaller critical separation.
For example, Cen \textit{et al.} report that the creation and
annihilation of oxygen vacancies at the surface could change the
polar structure of the LAO film and the electronic properties of
the interface~\cite{12}.

Our results also provide clear evidence of an asymmetry in the
electronic reconstruction at $n$-type and $p$-type interfaces.
Fig. 2a) and b) illustrate the $xy$-averaged spatial distribution
of the transferred electron and hole densities along the (001)
direction~\cite{32}. At the $n$-type interface, the electron
density decays monotonically into the STO, indicating a bound
state. However, the holes are not bound to the $p$-type interface
because their density maximum is located in the middle of the STO
substrate. (This feature persists as we thicken the STO.) This
suggests that holes diffuse into the STO substrate while electrons
stay close to the interface. Because carrier mobility depends on
trapping and scattering, the asymmetry in these distributions
provides a tentative explanation for the observed difference in
conductivity at these two interfaces. Under an applied field in
the $xy$-plane, electrons scatter only from defects within a few
nanometers of the interface but the holes scatter off defects
throughout the entire substrate. Given the high quality of the
interface [9], electrons should be much more mobile than the
holes.  Furthermore, holes will easily be trapped on inevitable
oxygen vacancies in the substrate which will further reduce their
mobility.

There is a single experiment that infers the presence of
$32\pm6\%$ oxygen vacancies per interfacial unit cell and
attributes the insulating behavior of the $p$-type interface to
the vacancies~\cite{9}. However, our DFT calculations show that
oxygen vacancies, just like holes, prefer to stay away from the
$p$-type interface~\cite{33}.  Thus simulating a clean $p$-type
interface is certainly consistent theoretically. The discrepancy
between experiment and theory needs to be resolved by additional
experiments and theoretical calculations. Nevertheless, we agree
that oxygen vacancies tend to trap and immobilize the hole
carriers and thus render the $p$-type interface insulating.

Then this begs the question: why are the electrons bound while the
holes are not? Our answer hinges on topological asymmetry. The
holes reside on O $p$-orbitals and the oxygen atoms form a
continuous network across the LAO/STO interface (Fig. 2d). We have
calculated the hopping matrix elements between neighboring O
$p$-orbitals and found no significant discontinuity at the
interface. Hence, the voltage across the polar LAO drives holes
into the STO, but once there, they delocalize into the substrate.

However, the situation is very different for electrons. In STO and
LAO, the conduction band edge states reside on the Ti and La
cations, respectively. The Ti occupies the $B$ site in STO while
in LAO the La occupies the $A$ site (in the standard $ABO_3$
nomenclature for perovskites), resulting in a discontinuity in the
arrangement of the electronically active transition metals across
the $n$-type interface (Fig. 2c). A detailed examination shows
that the electrons occupy mainly the Ti $d_{xy}$-orbitals but have
a small weight on the La $d_{xy}$-orbitals in the first LaO layer
at the interface. We calculate that the interfacial hopping matrix
element $t$ between Ti $d_{xy}$-orbitals in the TiO$_2$ plane and
the La $d_{xy}$-orbitals in the LaO layer (Fig. 2c), which does
not exist in either bulk oxide and is unique to the interface, is
0.7 eV, about 100 times larger than that between neighboring Ti
$d_{xy}$-orbitals in STO. Two factors make this hopping so large:
first, La 5$d$-orbitals are spatially more extended than Ti
3$d$-orbitals, and second, the distance between the La and Ti is
$86\% (\sqrt{3}/2)$ of the Ti-Ti distance in bulk STO.

Thus, as a zeroth order approximation, we focus only on the
interacting Ti and La at the interface. The strong interfacial
hopping causes the Ti $d_{xy}$ state to lower its energy and the
La $d_{xy}$ energy to rise. In chemical language, we form a pair
of bonding and anti-bonding states between the two cations across
the interface. Therefore, compared to the other Ti $d$-orbitals in
the substrate, the Ti $d$-orbitals at the $n$-type interface are
more energetically favorable and bind electrons to the interface.

First principle simulations provide complete control over the
system and allow us to verify the above picture via a gedanken
experiment. We take the optimized (relaxed) $n$-type interface
structure as the reference. We manually move only the Ti atom at
the interface by an amount $z$, allow the electronic system to
achieve self-consistency, and calculate the new interfacial
properties including the hopping matrix element $t$. When $z>0$,
the Ti atom moves away from the interface, the Ti-La bond is
elongated and $t$ is weakened. When $z<0$, the Ti atom moves
towards the interface, shortens the La-Ti bond and strengthens
$t$. Fig. 3 shows how the transferred electron density responds to
changing $z$. As the Ti moves toward the interface, the electron
density further localizes on the first Ti atom at the interface.
When the Ti moves away, the biggest peak in the density profile
then appears on the next Ti atom away from the interface. These
results provide strong direct evidence that the large interfacial
hopping between Ti $d$- and La $d$-orbitals traps and binds the
electrons close to the interface.

In conclusion, we reveal a fundamental asymmetry in the electronic
reconstruction at the $n$-type and $p$-type LAO/STO interface,
which is accounted for by a strongly enhanced electronic hopping
between the interfacial cations that host the carriers. This
mechanism opens a pathway to nanoengineering electronic states at
heterointerfaces. For example, doping different cations at the
$n$-type interface would change the spatial extent of atomic
orbitals, modify the hopping element and thus tune the binding of
the electrons to the interface. Mechanically, cation motions
relative to the interface, e.g. by a ferroelectric or
pieozoelectric effect, could similarly modify the distribution of
the interfacial electron states.

We are grateful to Charles H. Ahn, Frederick J. Walker, Victor E.
Henrich, Eric I. Altman and John C. Tully for stimulating
discussions. This work was supported by the National Science
Foundation under Contract No. MRSEC DMR 0520495. The Bulldog
parallel clusters of the Yale High Performance Computing center
provided computational resources.

\newpage

\begin{figure}[h]
\includegraphics[angle=0,width=15cm]{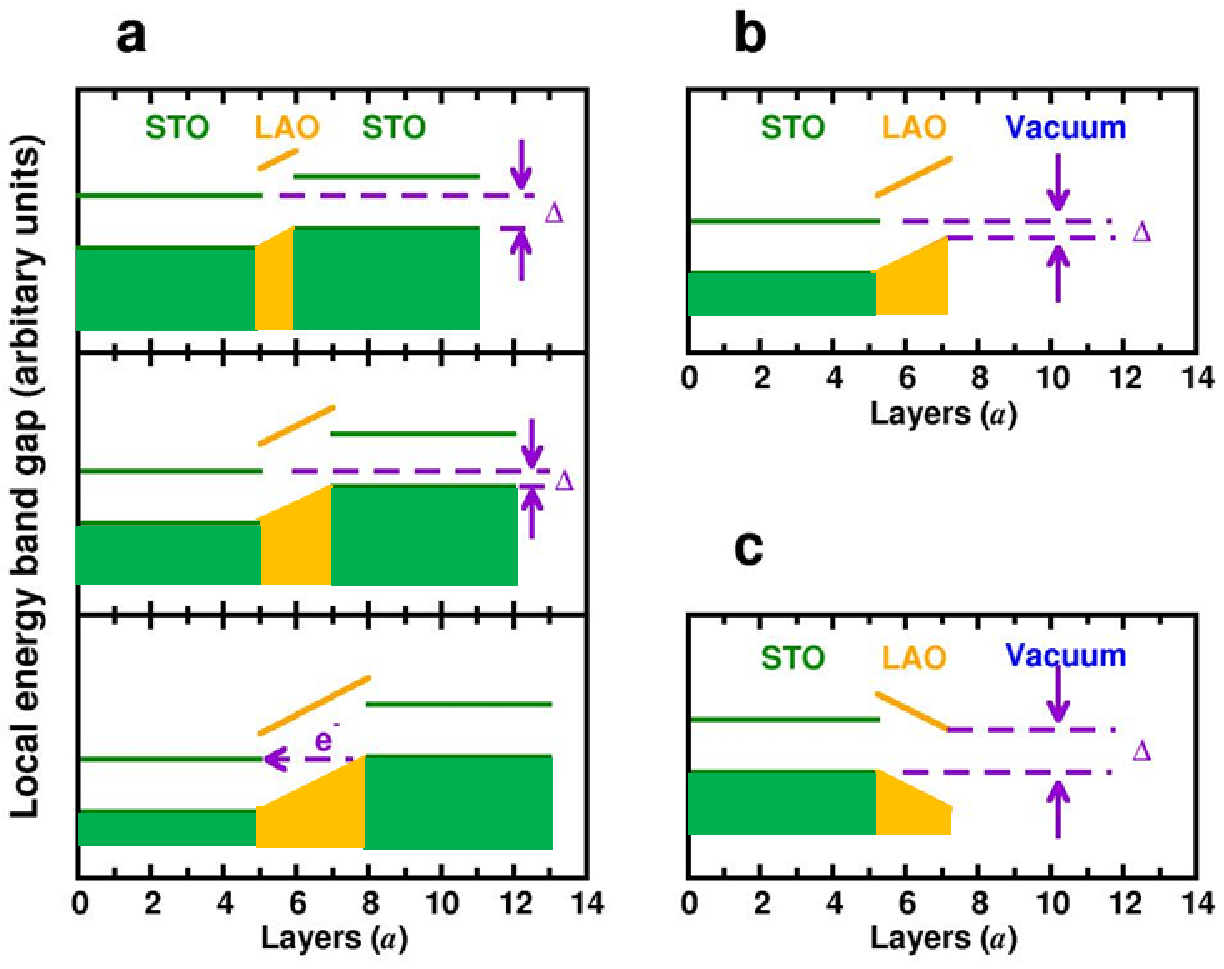}
\label{fig1} \caption{ (Color online) Schematic local energy gaps
of a) \textit{np}-type, b) \textit{n}-type and c) \textit{p}-type
interfaces, respectively. For b) and c), the righthand LAO
surfaces face the vacuum. The shaded parts are filled valence
bands and the upper parts are empty conduction bands. In a) from
top to bottom, the energy gap of the interface system $\Delta$
(highlighted in dashed purple/dark gray) is diminished by adding
LAO layers until it disappears. For \textit{p}-type interfaces,
the energy gap is not reduced until the conduction band edge of
LAO becomes lower than that of STO. $a$ is the STO lattice
constant.}
\end{figure}

\begin{table}[h]
\caption{DFT-LDA computed energy gaps and potential changes versus
the number of LAO layers $i$ for $np$-type, $n$-type and $p$-type
interfaces ($i$=0 refers to a pure STO substrate).  $\Delta$ is
the energy gap of the interface system with the corresponding
number of LAO layers. $V_i$ is the macroscopic potential change
due to adding the $i$-th LAO layer.}
\begin{center}
\begin{tabular}{ccc|ccc|ccc}
\hline
\hline & $np$-type& & &$n$-type& & &$p$-type& \\
\hline $i$ & $\Delta$(eV) & $V_i$(eV) &  $i$  & $\Delta$(eV) &
$V_i$(eV) & $i$ & $\Delta$(eV) & $V_i$(eV)\\
\hline 0 & 1.85     &      & 0 & 1.85     &      & 0   & 1.85    &  \\
\hline 1 & 1.25     & 0.60 & 1 & 1.71     & 0.14 & 1   & 1.79    & 0.61 \\
\hline 2 & 0.55     & 0.70 & 2 & 1.28     & 0.43 & 2   & 1.82    & 0.72 \\
\hline 3 & metallic &      & 3 & 0.57     & 0.71 & 3   & 1.18    & 0.72 \\
\hline   &          &      & 4 & metallic &      & 4   & 0.49    & 0.69 \\
\hline   &          &      &   &          &      & 5   &
metallic&\\
\hline \hline
\end{tabular}
\end{center}
\end{table}

\begin{figure}[h]
\includegraphics[angle=-90,width=15cm]{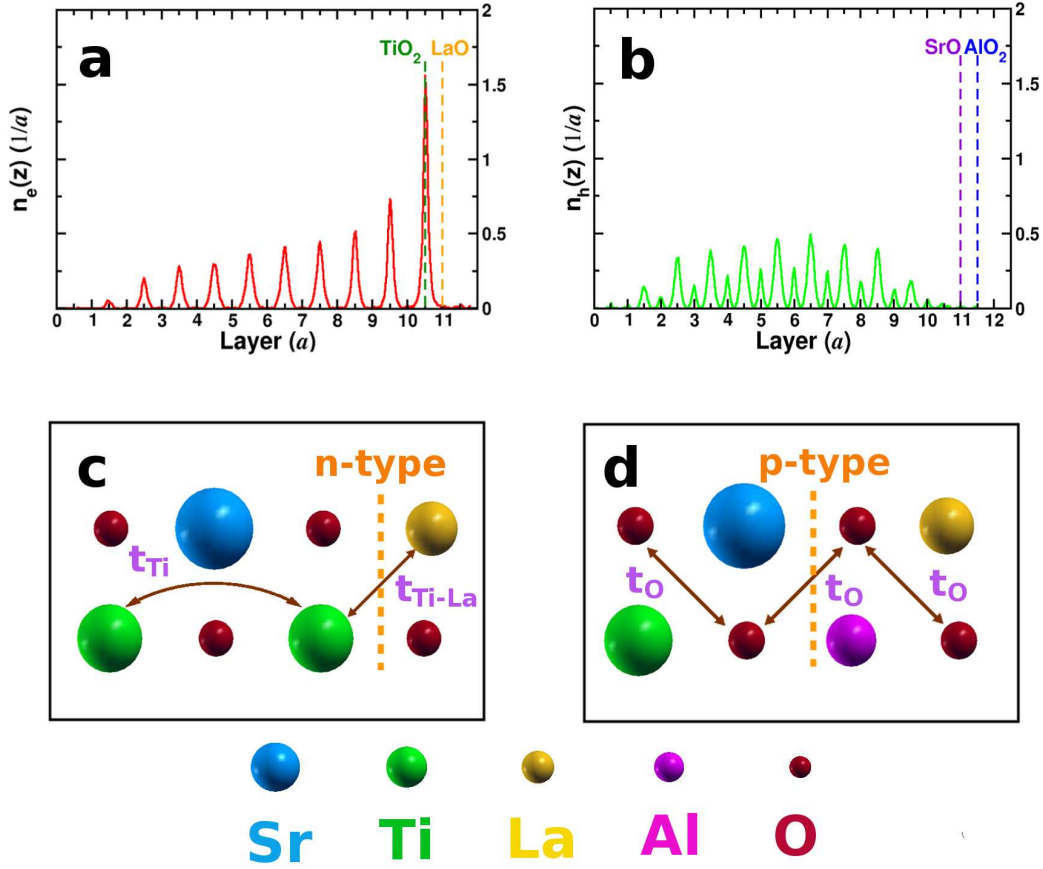}
\label{fig2} \caption{ (Color online) a) The total transferred
electron density integrated over the $xy$-plane in the STO
conduction bands at the $n$-type interface. The electrons are
bound to the $n$-type interface and decay into the STO substrate.
b) The total transferred hole density integrated over the
$xy$-plane in the STO valence bands at the $p$-type interface. The
holes are delocalized with the maximum around the center of STO
slab. Interfaces are highlighted by the colored dashed lines.
Layers are measured in units of the STO lattice constant $a$. The
integrals of these charge densities over the STO region are
normalized to unity. c) Schematic of different types of hopping
between neighboring Ti $d$-orbitals and between Ti $d$- and La
$d$-orbitals at the $n$-type interface. d) Schematic of hopping
between different O $p$-orbitals at the $p$-type interface. }
\end{figure}

\begin{figure}[h]
\includegraphics[angle=-90,width=15cm]{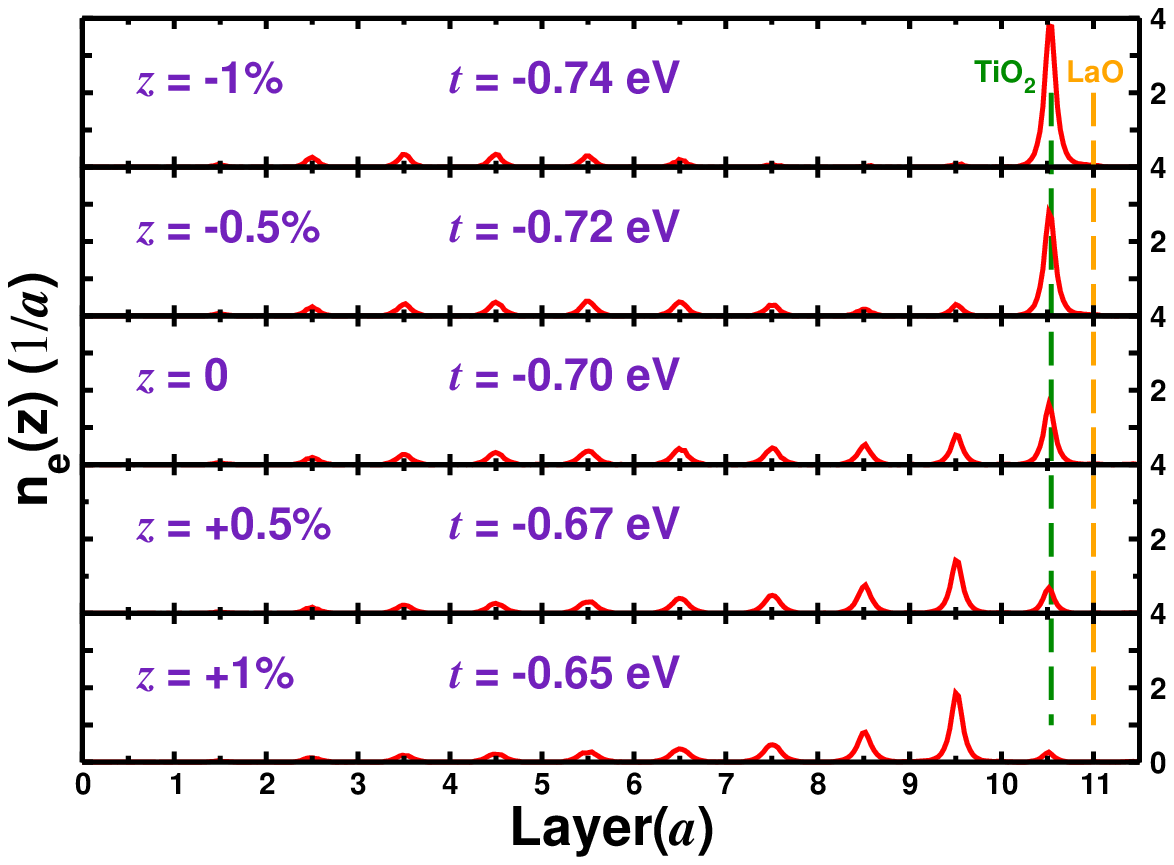}
\label{fig3} \caption{ Evolution of the total transferred
$xy$-integrated electron density in the STO slab and the
interfacial hopping matrix element $t$ as a function of the
movement of the Ti atom at the $n$-type interface. $z>0$: Ti moves
away from the interface; $z<0$: Ti moves towards the interface.
The integral of each electron density over the STO region is
normalized to unity. (Distances and the movement $z$ are measured
in units of the STO lattice constant $a$.)}
\end{figure}


\begin{thebibliography}{99}
\bibitem{1} S. B. Ogale, \textit{Thin Films and Heterostructures for Oxide Electronics} (Springer Verlag, Berlin, 2005)
\bibitem{2} E. Dagotto, Science \textbf{318}, 1076 (2007)
\bibitem{3} A. Ohtomo, D. A. Muller, J. L. Grazul and H. Y. Hwang, Nature \textbf{419}, 378 (2002)
\bibitem{4} A. Ohtomo and H. Y. Hwang, Nature \textbf{427}, 423 (2004)
\bibitem{5} S. Thiel, G. Hammerl, A. Schmehl, C. W. Schneider, J. Mannhart, Science \textbf{313}, 1942 (2006)
\bibitem{6} H. Y. Hwang, Science \textbf{313} 1895 (2006)
\bibitem{7} A. Brinkman, M. Huijben, M. Van Zalk, J. Huijben, U. Zeitler, J. C. Maan, W. G. Van der Wiel,
G. Rijnders, D. H. A. Blank and H. Hilgenkamp, Nature Mater.
\textbf{6}, 493 (2007)
\bibitem{8} N. Reyren, S. Thiel, A. D. Caviglia, L. Fitting Kourkoutis, G. Hammerl,
C. Ricther, C. W. Schneider, T. Kopp, A. S. R\"{u}etschi, D.
Jaccard, M. Gabay, D. A. Muller, J. M. Triscone, J. Mannhart,
Science \textbf{317}, 1196 (2007)
\bibitem{9} N. Nakagawa, H. Y. Hwang and D. A. Muller, Nature Mater. \textbf{5}, 204 (2006)
\bibitem{10} M. Huijben, G. Rijnders, D. H. A. Blank, S. Bals, S. Van Aert, J. Verbeeck,
G. Van Tendeloo, A. Brinkman and H. Hilgenkamp, Nature Mater.
\textbf{5}, 556 (2006)
\bibitem{11} P. R. Willmott, S. A. Pauli, R. Herger, C. M. Schlep\"{u}tz, D. Martoccia,
B. D. Patterson, B. Delley, R. Clarke, D. Kumah, C. Cionca and Y.
Yacoby, Phys. Rev. Lett. \textbf{99}, 155502 (2007)
\bibitem{12} C. Cen, S. Thiel, G. Hammerl, C. W. Schneider, K. E. Andersen, C. S. Hellberg,
J. Mannhart and J. Levy, Nature Mater. \textbf{7}, 298 (2008)
\bibitem{13} R. Pentcheva and W. E. Pickett, Phys. Rev. B \textbf{74}, 035112 (2006)
\bibitem{14} M. S. Park, S. H. Rhim and A. J. Freeman, Phys. Rev. B \textbf{74}, 205416 (2006)
\bibitem{15} J. M. Albina, M. Mrovec, B.Meyer, and C. Els\"{a}sser, Phys. Rev. B \textbf{76}, 165103 (2007)
\bibitem{16} S. Gemming, G. Seifert Acta Mater. \textbf{54}, 4299 (2006)
\bibitem{17} K. Janicka, J. P. Velev and E. Y. Tsymbal, J. Appl. Phys. \textbf{103}, 07B508 (2008)
\bibitem{18} Z. Zhong and P. J. Kelly, Europhys. Lett. \textbf{84} 27001 (2008)
\bibitem{19} U. Schwingenschl\"{o}gl and C. Schuster, Europhys. Lett. \textbf{81}, 17007 (2008)
\bibitem{20} Jaekwang Lee and Alexander A. Demkov, Phys. Rev. B \textbf{78}, 193104 (2008)
\bibitem{21} R. Pentcheva and W. E. Pickett, Phys. Rev. B \textbf{78}, 205106 (2008)
\bibitem{22} A. Kalabukhov, R. Gunnarsson, J. B\"{o}rjesson, E. Olsson, T. Claeson and D. Winkler,
Phys. Rev. B \textbf{75}, 121404(R) (2007)
\bibitem{23} G. Herranz, M. Basleti\'{c}, M. Bibes, C. Carr\'{e}t\'{e}ro, E. Tafra, E. Jacquet, K. Bouzehouane,
C. Deranlot, A. Hamzi\'{c}, J. M. Broto, A. Barth\'{e}l\'{e}my,
and A. Fert, Phys. Rev. Lett. \textbf{98}, 216803 (2007)
\bibitem{24} W. Siemons, G. Koster, H. Yamamoto, W. A. Harrison, G. Lucovsky, T. H. Geballe,
D. H. A. Blank, and M. R. Beasley, Phys. Rev. Lett. \textbf{98},
196802 (2007)
\bibitem{25} J. N. Eckstein, Nature Mater. \textbf{6}, 473 (2007)
\bibitem{26} S. Okamoto and A. J. Millis, Nature \textbf{428}, 630 (2004)
\bibitem{27} M. C. Payne, M. P. Teter, D. C. Allan, T. A. Arias, J. D. Joannopoulos, Rev. Mod. Phys. \textbf{64}, 1045 (1992)
\bibitem{28} W. Kohn and L. J. Sham, Phys. Rev. \textbf{140}, A1133 (1965)
\bibitem{29} D. Vanderbilt, Phys. Rev. B \textbf{41} 7892 (1990)
\bibitem{30} The STO slabs in our simulations are
always terminated in a SrO plane to avoid surface effects, as
described in the EPAPS Document No. XXX.
\bibitem{31} The reason that the critical separation for $n$-type interfacesis larger than that
for $np$-type is that the potential change by the first two unit
cells of LAO (0.14 and 0.43 eV) are well below the extrapolated
value 0.7 eV (See Table 1).
\bibitem{32} To be clear, the electron density (for $n$-type interface) in Fig. 2a is the
weighted sum of all the occupied states in the STO conduction
bands and the hole density (for $p$-type interface) in Fig. 2b is
the weighted sum of all the unoccupied states in the STO valence
bands.
\bibitem{33} See the EPAPS Document No. XXX for details.
\end{thebibliography}
\end{document}